\documentstyle[sprocl,psfig]{article}

\def\be{\begin{equation}}
\def\ee{\end{equation}}
\def\bea{\begin{eqnarray}}
\def\eea{\end{eqnarray}}
\newcommand\bfg[1]{\begin{figure}\vspace{#1cm}}
\newcommand\efg{\end{figure}}

\def\nl{n_{\mathrm liquid}}

\def\n0{n_{\mathrm outside}}

\def\mathrm{ }   

\begin{document}
\title{SONOLUMINESCENCE AND THE QED VACUUM~\footnote{ Presented by 
S.~Liberati. To appear in the Proceedings of the
``Fourth Workshop on Quantum Field Theory under the Influence of
External Conditions''.}}

\author{S. LIBERATI}

\address{International School for Advanced Studies,\\ 
Via Beirut 2-4, 34014 Trieste, Italy\\
INFN sezione di Trieste\\E-mail: liberati@sissa.it} 

\author{MATT VISSER}

\address{Physics Department, Washington University,\\ 
Saint Louis MO 63130-4899, USA\\E-mail: visser@kiwi.wustl.edu}

\author{F. BELGIORNO}

\address{Universit\`a degli studi di Milano, Dipartimento di Fisica,\\
Via Celoria 16, 20133 Milano, Italy\\E-mail: belgiorno@mi.infn.it}

\author{D.W. SCIAMA}

\address{International School for Advanced Studies,\\ 
Via Beirut 2-4, 34014 Trieste, Italy\\
International Center for Theoretical Physics, \\
Strada Costiera 11, 34014 Trieste, Italy\\
Physics Department, Oxford University, Oxford, England\\
E-mail: sciama@sissa.it}

\author{SISSA-ref 131/98/A}


\maketitle

\abstracts{
In this talk I shall describe an extension of the quantum-vacuum
approach to sonoluminescence proposed several years ago by
J.~Schwinger. We shall first consider a model calculation
based on Bogolubov coefficients relating the QED vacuum in the
presence of an expanded bubble to that in the presence of a collapsed
bubble. In this way we shall derive an estimate for the spectrum
and total energy emitted. This latter will be shown to be proportional
to the volume of space over which the refractive index changes, as
Schwinger predicted. After this preliminary check we shall deal
with the {\em physical constraints} that any viable dynamical model
for SL has to satisfy in order to fit the experimental data. We
shall emphasize the importance of the timescale of the change in 
refractive index.  This discussion will led us to propose a
somewhat different version of dynamical Casimir effect in which
the change in volume of the bubble is no longer the only source
for the change in the refractive index.}

\section{Introduction}

Sonoluminescence (SL) is the phenomenon of light emission by a
sound-driven gas bubble in fluid~\cite{Physics-Reports}. The
intensity of a standing sound wave can be increased until the
pulsations of a bubble of gas trapped at a velocity node have
sufficient amplitude to emit picosecond flashes of light.  The
basic mechanism of light production is still controversial. We
shall start by presenting a brief summary of the main
experimental data (as currently understood) and their sensitivities
to external and internal conditions. The most common situation is
that of an air bubble in water. SL experiments usually deal with
bubbles of ambient radius $R_{\mathrm ambient} \approx 4.5 \; \mu
{\rm m}$. The bubble is driven by a sound wave of frequency of $20
- 30$ kHz.  During the expansion phase, the bubble radius reaches
a maximum of order $R_{\mathrm max}\approx 45 \; \mu {\rm m}$,
followed by a rapid collapse down to a minimum radius of order
$R_{\mathrm max}\approx 0.5 \; \mu {\rm m}$. The photons emitted
by such a pulsating bubble have typical wavelengths of the order
of visible light. The minimum observed wavelengths range between
$200 \;{\rm nm}$ and $100 \; {\rm nm}$. This light appears distributed
with a power-law spectrum (with exponent depending on the noble
gas admixture entrained in the bubble) with a cutoff in the extreme
ultraviolet. If one fits the data to a Planck black-body spectrum
the corresponding temperature is several tens of thousands of Kelvin
(typically $70,000\; {\rm K}$, though estimates varying from $40,000
\; {\rm K}$ to $100,000 \; {\rm K}$ are common). There is considerable
doubt as to whether or not this temperature parameter corresponds
to any real physical temperature.  There are about one million
photons emitted per flash, and the average total power released is
$30 \;\hbox{ mW} \leq W \leq 100 \;\hbox{ mW}$.  The photons appear
to be created in a very tiny spatio-temporal region: of order
$10^{-1} \mu {\rm m}$ and on timescales $\tau \leq 50 \;\hbox{\rm
ps}$ (there have been claims that flash duration is less than $100\;
{\rm fs}$, though more recent claims place flash duration in the
range $50 - 250$ ps~\cite{Flash1,Flash2}).  A truly successful
theory of SL must also explain a whole series of characteristic
sensitivities to different external and internal conditions. Among
these dependencies the main one is surely the mysterious catalytic
role of noble gas admixtures.  Other external conditions that
influence SL are magnetic fields and the temperature of the water
(see~\cite{Physics-Reports}).

These are only the most salient experimental dependencies of
the SL phenomenon.  In explaining such detailed and specific
behaviour the Casimir approach (the QED vacuum approach) encounters
the same problems as other approaches have. Nevertheless we shall
argue that SL explanations using a Casimir-like framework are
viable.

\subsection{Quasi-static Casimir models: Schwinger's 
approach}\label{subsec:schw}

The idea of a ``Casimir route'' to SL is due to Schwinger who
several years ago wrote a series of
papers~\cite{Sc1,Sc2,Sc3,Sc4,Sc5,Sc6,Sc7} regarding the so-called
dynamical Casimir effect. Considerable confusion has been caused
by Schwinger's choice of the phrase ``dynamical Casimir effect''
to describe his model. In fact, his original model is not
dynamical and is at best quasi-static as the heart of the model
lies in comparing two static Casimir energy calculations: that for
an expanded bubble with that for a collapsed bubble.  Schwinger
estimated the energy emitted during this collapse as being
approximately equal to the change in the static Casimir energy.
The static Casimir energy of a dielectric bubble (of dielectric
constant $\epsilon_{{\mathrm  inside}}$) in a dielectric background
(of dielectric constant $\epsilon_{{\mathrm  outside}}$) is
\begin{equation}
E_{cavity}
=+\frac{1}{6 \pi} \hbar c R^3 K^4
\left(
{1\over\sqrt{\epsilon_{\mathrm {\mathrm  inside}}}}-
{1\over\sqrt{\epsilon_{\mathrm {\mathrm  outside}}}}
\right) +\cdots.
\label{Esch}
\end{equation}
Here the dots stand for additional sub-dominant finite volume
effects~\cite{CMMV1,CMMV2,MV}. The quantity $K$ is a high-wavenumber
cutoff that characterizes  the wavenumber at which the real parts
of the refractive indices drop to their vacuum  values. Hence $K$
is a physical cutoff given by condensed matter physics, not a
regularization parameter to be renormalized away. This
cutoff can be interpreted as the typical length scale beyond
which the notion of a continuous dielectric medium is no longer
meaningful. The result (\ref{Esch}) can also be rephrased in the 
clearer and more general form  as~\cite{CMMV1,CMMV2,MV}: 
\begin{equation}
E_{ cavity} = + 2 V \int \frac{d^3\vec{k}}{(2 \pi)^3} \; \frac{1}{2}
\hbar  \left[ \omega_{\mathrm {\mathrm  inside}}(k) - \omega_{\mathrm
{\mathrm  outside}}(k)  \right] +
\cdots
\end{equation}
where it is evident that the Casimir energy can be interpreted as
a difference in zero point energies due to the different dispersion
relations inside and outside the bubble. In the case of SL
$\omega_{\mathrm inside}(k)\approx ck$, $\omega_ {\mathrm
outside}(k)=ck/n$ for $k < K$, and $\omega_ {\mathrm outside}(k)\approx
ck$ for $k > K$. In Schwinger's original model he took $n\approx\nl\approx
1.33$, $R\approx R_{\mathrm max}$,  and $K\approx 2\pi/400 \;{\rm
nm^{-1}}$, leading to about three million emitted photons~\cite{Sc4}.

The three main strengths of models based on zero point fluctuations
are:\\
1) The vacuum production of photon pairs allows
for the very short timescales that one requires to fit data. Typically
one expects these timescales to be of the order of the time that the
zero point modes of the EM field takes to be correlated on the bubble
scale.  (Roughly the light-crossing time for the bubble.) For a bubble
of radius $0.5$ microns this time scale is about $1.6$ femtoseconds,
which is certainly sufficiently rapid to be compatible with observed
flash duration.\\ 
2) One does {\em not} need to achieve ``real'' temperatures of thousands of
Kelvin inside the bubble. Quasi-thermal behaviour is generated in
quantum vacuum models by the squeezed nature of the two photon states
created, and the ``temperature'' parameter is a
measure of the squeezing, not a measure of any real physical
temperature~\footnote{This ``false thermality'' must not be
confused with the very specific phenomenon of Unruh temperature. 
In that case, valid only for uniformly accelerated observers in flat
space, the temperature is related to the constant value of the
acceleration.  Instead, in the case of squeezed states, the apparent
temperature can be related to the degree of squeezing~\cite{BLVS} of
the real photon pairs generated via the dynamical Casimir effect.}.\\
3) There is no actual production of far ultraviolet photons (because
the refractive index goes to unity in the far ultraviolet) so one does
not expect the dissociation effects in water that other models
imply. Models based on the quantum vacuum automatically provide a
cutoff in the far ultraviolet from the behaviour of the refractive
index. Moreover this cutoff appears to be sensitive to the water
temperature in such a way to explain the former described experimental
dependencies---this observation going back to Schwinger's first papers
on the subject~\cite{Sc4}.

Thus one key issue in Schwinger's model is simply that of
calculating static Casimir energies for dielectric spheres.  It
must be stressed that there is still considerable disagreement on
this calculation.  Milton, and Milton and
Ng~\cite{KM} strongly criticize Schwinger's result.  These points
have been discussed extensively in~\cite{CMMV1,CMMV2,MV} where it is
emphasized that one has to compare two different geometrical
configurations, and different quantum states, of the same spacetime
regions. In a situation like Schwinger's model for SL one has to
subtract from the zero point energy (ZPE) for a vacuum bubble in water
the ZPE for water filling all space. It is clear that in this case the
bulk term is physical and {\em must} be taken into account. In
the situation pertinent to sonoluminescence, the total volume
occupied by the gas is not at all conserved (the gas is truly
compressed), and it is far too naive to simply view the ingoing
water as flowing coherently from infinity (leaving voids
filled with air or vacuum somewhere in the apparatus). Since
the density of water is approximately but not exactly constant, the
influx of water will instead generate an outgoing density wave which
will be rapidly damped by the viscosity of the fluid. The few phonons
generated in this way are surely negligible.

\subsection{Eberlein's dynamical model for SL}

The quantum-vacuum approach to SL was extended in the work
of Eberlein~\cite{Eberlein}.  The basic mechanism in Eberlein's
approach is a dynamical Casimir effect:  Photons are produced due
to a change of the refractive index in the portion of space between
the minimum and the maximal bubble radius (a related discussion
for time-varying refractive index is due to
Yablonovitch~\cite{Yablonovitch}).  This physical framework is
actually implemented via a boundary between two dielectric media
which accelerates with respect to the rest frame of the quantum
vacuum state. The adiabatic change in the zero-point modes of the
fields reflects in a non-zero radiation flux.

In the Eberlein analysis the motion of the bubble boundary is taken  
into account by introducing a velocity-dependent perturbation to the
usual EM Hamiltonian:
\begin{eqnarray}
H_{\epsilon} &\!=&\!
\frac{1}{2} \int{\rm d}^3{\bf r}
\left(
{{\bf D}^2\over\epsilon} + {\bf B}^2
\right)\;,\:\:\:
\Delta H \!=\!
\beta
\int{\rm d}^3{\bf r}
\frac{\epsilon -1}{\epsilon}\;
({\bf D}\wedge{\bf B})\cdot{\bf\hat r}\;.
\end{eqnarray}
This is an approximate low-velocity result coming from a power
series expansion in the speed of the bubble wall $\beta= \dot R/c$.
(The bubble wall is known to collapse with supersonic velocity,
values of Mach 4 are often quoted, but this is still
completely non-relativistic with $\beta\approx 10^{-5}$.) 

Eberlein's final result for the energy radiated over one acoustic   
cycle is:
\begin{equation}
{\cal W} = 1.16\:\frac{(n^2-1)^2}{n^2}\,\frac{1}{480\pi}
\left[{\hbar\over c^3}\right]
\int_{0}^{T} {\rm d}\tau\; \frac{\partial^5 R^2(\tau)}{\partial
        \tau^5}\,R(\tau)\beta(\tau)\;.
\end{equation}
(Eberlein approximates $n_{\mathrm {\mathrm inside}} \approx
n_{\mathrm air} \approx 1$ and sets $n_{\mathrm {\mathrm outside}}
= n_{\mathrm water} \to n$.  The $1.16$ is the result of an
integral is estimated numerically.) In this mechanism the massive
burst of photons is produced at and near the turn-around at the
minimum radius of the bubble. There the velocity rapidly changes
sign, from collapse to re-expansion. This means that the acceleration
is peaked at this moment, and so are higher derivatives of the
velocity.

The main points of strength of the Eberlein model are the same as
previously listed for the Schwinger model. However, Eberlein's
model exhibits a significant weakness (which does not apply to the
Schwinger model):\\
The calculation is based on an adiabatic approximation which does
not seem consistent with results~\footnote{The adiabatic approximation
is actually justified in the case of a model based on the bubble 
collapse case by the fact that the frequency $\Omega$ of the driving 
sound (and hence the timescale of the bubble collapse) is of the order 
of tens of kHz, while that of the emitted light is of the order of 
$10^{16}$ Hz. The problem we stress here is instead related to the 
``self-consistency'' of Eberlein's model.}. In order to fit the 
experimental values the model requires, as an external input, the 
bubble radius time dependence.  This is expressed as a function of a 
parameter $\gamma$ which describes the time scale of the collapse and
re-expansion process.  In order to fit the experimental values for
$\cal W$ one has to fix $\gamma \approx 10 \;{\rm fs}$.  This is
far too short a time to be compatible with the adiabatic
approximation.  Although one might claim that this number
can ultimately be modified by the eventual inclusion of
resonances it would seem reasonable to take this ten femtosecond
figure as a first self-consistent approximation for the characteristic
timescale of the driving system (the pulsating bubble).  Unfortunately,
the characteristic timescale of the collapsing bubble then comes
out to be of the same order of the characteristic period of the
emitted photons. This shows that attempts at bootstrapping the
calculation into self-consistency instead bring it to a regime
where the adiabatic approximation underlying the scheme cannot be
trusted.

This discussion has lead us to discover a quite intricate
situation. We have on the one hand simple estimates of the
vacuum energy that can be involved in SL, estimates that are still
the object of heated debate, and on the other hand we have a
dynamical approach to the problem that seems to be partially self
contradictory.  In order to resolve the first issue and to understand
the proper framework to deal with the second we shall now consider
what we can best view as a ``toy model''. In spite of its simplicity
this toy model will allow us to capture some basic results that
we hope will guide future research on the ``Casimir route'' to
sonoluminescence.
\section{Bogolubov approach on a single oscillation}
Let us consider a single pulsation of the bubble.  At this  stage
of development, we are not concerned with the dynamics of the bubble
surface.  In analogy with the subtraction procedure of the static
calculations of Schwinger~\cite{Sc1,Sc2,Sc3,Sc4,Sc5,Sc6,Sc7} or of
Carlson {\em et al.}~\cite{CMMV1,CMMV2,MV} we shall consider two
different configurations of space.  An ``in'' configuration with
a bubble of dielectric constant $\epsilon_{\mathrm {\mathrm
inside}}$ (typically vacuum) in a medium of dielectric constant
$\epsilon_{\mathrm {\mathrm outside}}$, and an ``out'' one in which
one has just the latter medium (dielectric constant $\epsilon_{\mathrm
{\mathrm outside}}$) filling all space.  These two  configurations
will correspond to two different bases for the quantization of the
field. (For the sake of simplicity we take, as Schwinger did, only
the electric part of QED, reducing the problem to a type of
scalar electrodynamics).  The two bases will be related by Bogolubov
coefficients in the usual way. Once we determine these coefficients
we easily get the number of created particles per mode and from
this the spectrum. We shall also make a consistency check by a
direct confrontation between the change in static Casimir
energy and the sum, $E=\sum_{k} \omega_{k} n_{k}$, of the
energies of the emitted photons.

\subsection{Bogolubov coefficients}

We use the Schwinger framework. In spherical coordinates and with a  
time independent dielectric constant
\begin{equation}
\epsilon {\partial^2\over\partial t^2} E-\nabla^{2} E=0.
\end{equation}
Solutions are of the form
\begin{equation}
E=e^{i\omega t} {G(r)\over \sqrt{r}} Y_{lm}(\Omega).
\end{equation}
Then one finds
\begin{equation}
G^{''}+{1\over r}G^{'}+\left(\lambda^{2}-{(l+1/2)^2 \over r^{2}} \right)G=0.
\end{equation}
where $\lambda^2 = \epsilon \omega^2$.  This is the standard Bessel
equation, it admits as solutions Bessel functions of the first
kind, $J_{\nu}(\lambda r)$, and  Neumann functions, $N_{\nu}(\lambda
r)$ (Bessel functions of the second kind), with $\nu=l+1/2$.

For the ``in'' QED vacuum we have to take into account that the
dielectric constant changes at the bubble wall. In fact we have
\begin{equation}
\epsilon=\left \{
\begin{array}{llll}
\epsilon_{1} & = & \mbox{dielectric constant of air} & \mbox{if
$r< R$},\\
\epsilon_{2} & = & \mbox{dielectric constant of water} & \mbox{if $r
>R$.}
\end{array}
\right.
\end{equation}
We now use the fact that the dielectric constant of air is
approximately equal 1 and shall deal only with the constant of
water ($n= \sqrt{\epsilon_{2}} \approx1.3$)
For the eigenmodes of the ``in'' state one has 
\begin{equation}
G^{\mathrm  in}_{\nu}(n,\omega,r)=\left \{
\begin{array}{ll}
A_{\nu} J_{\nu}(\omega_{\mathrm  in} r) & \mbox{if $r< R$,}\\
B_{\nu} J_{\nu}(n \omega_{\mathrm  in} r)+C_{\nu} N_{\nu}(n \omega_{\mathrm  in} r)&
\mbox{if $r > R$.}
\end{array}
\right.
\end{equation}
The coefficients $A_{\nu}$, $B_{\nu}$ and $C_{\nu}$ are determined by
the matching conditions
\begin{equation}
\begin{array}{lll}
A_{\nu} J_{\nu}(\omega_{\mathrm  in}  R)&=&
B_{\nu} J_{\nu}(n \omega_{\mathrm  in} R)+C_{\nu}N_{\nu}(n \omega_{\mathrm  in} R),\\
A_{\lambda} J_{\nu}{'}(\omega_{\mathrm  in} R)&=&
B_{\nu} J_{\nu}{'}(n \omega_{\mathrm  in} R)+C_{\nu}
N_{\nu}{'}(n \omega_{\mathrm  in} R).
\end{array}
\label{coef}
\end{equation}   
The eigenmodes for the ``out'' QED vacuum are easily obtained solving
the same equation but for a space filled with an homogeneous
dielectric~\footnote{
Keeping $R_{\mathrm min}$ finite
significantly complicates the calculation but does not give much more 
physical information.}. 
\begin{equation}
G^{\mathrm out}_{\nu}(n,\omega_{\mathrm out},r)=J_{\nu}(n
\omega_{\mathrm out} r).
\end{equation}
The Bogolubov coefficients are defined as
\begin{eqnarray}  
\alpha_{ij}
&=&
({E_{i}^{\mathrm  out}},{E_{j}^{\mathrm in}}),\:\:\: 
\beta_{ij}=({E_{i}^{\mathrm  out}}^{*}, {E_{j}^{\mathrm in}}) 
\end{eqnarray}
where the naive scalar product is as usual~\footnote{There
are subtleties in the definition of scalar product which we shall
deal with more fully in~\cite{letter,LBVS}. The naive scalar product
adopted here is good enough for a qualitative discussion.}
\begin{equation}
(\phi_{1},\phi_{2})=+i \int_{\Sigma}\phi_{1}^*
\stackrel{\leftrightarrow}{\partial}_{0}\phi_{2} \: d^{3}x.
\end{equation}
We are mainly interested in the coefficient $\beta$ since $|\beta|^{2}$
is linked to the total number of particles created.
By a direct substitution it is easy to find
\begin{eqnarray}
\beta
&=& (\omega_{\mathrm  in}-\omega_{\mathrm  out})
e^{i(\omega_{\mathrm  out}+\omega_{\mathrm  in})t} \delta_{l
l^{\prime}}\delta_{m,-m^{\prime}}
\nonumber\\
&& \qquad
\int_{0}^{\infty} G^{\mathrm  out}_{l}(n,\omega_{\mathrm  out},r)
\;
G^{\mathrm  in}_{l^{\prime}}(n,\omega_{\mathrm  in},r) 
\: r dr.
\label{beta1} 
\end{eqnarray}
After some work, the squared $\beta$ 
coefficients summed over $l$ and $m$ can be shown to be~\cite{letter,LBVS}
\begin{eqnarray}
\left|\beta(\omega_{\mathrm  in},\omega_{\mathrm  out})\right|^{2}
&=& 
\left(
\frac{n^2-1}{n^2}   
\frac{\omega_{\mathrm  in}^2 R}{\omega_{\mathrm  out}+\omega_{\mathrm  in}}
\right)^2
\sum_{\nu}(2\nu)
\left| A_{\nu} \right|^{2}
\nonumber\\
&&
\times
\left[
\frac{
W[J_{\nu}(n\omega_{\mathrm  out}r), J_{\nu}(\omega_{\mathrm  in}r)]_{R}}
{(n\omega_{\mathrm  out})^2-\omega_{\mathrm  in}^2} 
\right]^2.
\label{E-beta-squared}
\end{eqnarray}
The number spectrum and total energy content of the emitted photons are
\begin{eqnarray}
\label{E:E}
{dN(\omega_{\mathrm  out})\over d\omega_{\mathrm  out}}
&=&\left(\int|\beta(\omega_{\mathrm  in},\omega_{\mathrm  out})|^{2}
d\omega_{\mathrm  in} \right),\\
E&=&\hbar \int {dN(\omega_{\mathrm  out})\over d\omega_{\mathrm  out}} \;
\omega_{\mathrm  out} \; d\omega_{\mathrm  out}.
\end{eqnarray} 
These expressions are too complex to allow an analytical resolution of
the problem (except for the $R\to\infty$ limit).
 
\subsection{Large-volume analytic limit}
 
In this limit the total energy emitted should be approximately
equal to the leading contribution in $R$ of the Casimir energy in
the ``in'' state (a volume term if Schwinger was qualitatively
correct, a surface or curvature one otherwise).  Technically, if
$R$ is very large (but finite in order to avoid infra-red divergences)
then the ``in'' and the ``out'' modes can both be approximated by
ordinary Bessel functions: $G^{\mathrm in}(n,\omega,r) 
\approx J_{\nu}(\omega_{\mathrm  in} r)$, $G^{\mathrm  
out}(n,\omega,r)\approx J_{\nu}(n\omega_{\mathrm  out} r)$. 
The Bogolubov coefficients simplify
\begin{eqnarray}
\beta_{ij}
=({E^{\mathrm  out}_{i}}^*,E^{\mathrm  in}_{j})
\approx
\left(1-\frac{1}{n}\right) e^{i\omega_{\mathrm  in}(1/n +1)t}
\delta_{ll^{\prime}}\delta_{m,-m^{\prime}} 
\delta(\omega_{\mathrm  in}-n\omega_{\mathrm  out}).
\label{bogse}
\end{eqnarray}
This implies
\begin{eqnarray}
|\beta(\omega_{\mathrm  in},\omega_{\mathrm  out})|^{2}
&\approx&
\left(1-\frac{1}{n}\right)^{2}
\sum_{l}(2l+1){R\over2\pi c}\delta(\omega_{\mathrm in}
-n\omega_{\mathrm  out}), 
\end{eqnarray}
where we have invoked the standard scattering theory result that
$(\delta^{3}(k))^{2} = V \delta^{3}(k) /(2\pi)^3$, specialized to
the fact that we have a 1-dimensional delta function.  The summation
over angular momenta can be estimated as~\cite{LBVS}
\begin{eqnarray}
\sum_{l=0}^{l_{\mathrm  max}}(2l+1)
\approx 
l_{\mathrm  max}^{2}(\omega_{\mathrm  out})
&\approx& \left(R \, n \,
\frac{\omega_{\mathrm  out}}{c}\right)^2.
\end{eqnarray}
This finally gives
\begin{eqnarray}
|\beta(\omega_{\mathrm  in},\omega_{\mathrm  out})|^{2}&\approx 
&\left(n-1\right)^{2} {R^3\over2\pi c^{3}} \omega_{\mathrm  out}^{2} \;
\delta(\omega_{\mathrm  in}-n\;\omega_{\mathrm  out}).
\label{Ebog}
\end{eqnarray}
We can now compute the spectrum and the total energy of the emitted
photons
\begin{equation}
{dN(\omega_{\mathrm  out})\over d\omega_{\mathrm  out}}\approx   
 n^{2}
\; \left({n-1}\over{n}\right)^{2}
\; \frac{R^{3}\omega_{\mathrm  out}^{2}}{2\pi c^{3}}\;
\Theta(K-n \,\omega_{\mathrm  out}/c),
\end{equation}
and
\begin{equation}
E
\approx 
\frac{1}{8\pi\, n^{2}}
\; \left({n-1}\over{n}\right)^{2}   
\hbar c K \; (R K)^{3}.
\label{energy}
\end{equation}
Hence, inserting our results (\ref{Ebog}) into Eqs.~(\ref{E:E})
for $dN(\omega)/d\omega$ and $E$, we deduce a
spectrum that is proportional to phase space (and hence is a power
law), up to the cutoff frequency where $n\to 1$.  We interpret this
as definitive proof that indeed Schwinger was qualitatively right:  
The main contribution to the Casimir energy of a (large)
dielectric bubble is a bulk effect. The total energy radiated in
photons balances the change in the Casimir energy up to factors of
order one which the present analysis is too crude to detect. (For
infinite volume the whole calculation can be re-phrased in terms
of plane waves to accurately fix the last few prefactors.)

It is important to stress that Eq.~(\ref{Esch}) and Eq.~(\ref{energy})
are not identical (even if in the large $R$ limit the leading term
of Casimir energy of the ``in'' state and the total photon energy
coincide). One can easy see that the volume term we just found
[Eq.~(\ref{energy})] is of second order in $(n-1)$ and not of first
order like Eq.~(\ref{Esch}).  This is ultimately due to the fact
that the interaction term responsible for converting the initial
energy in photons is a pairwise squeezing operator~\cite{BLVS}.
Equation (\ref{energy}) demonstrates that any argument that attempts
to deny the relevance of volume terms to sonoluminescence due to
their dependence on $(n-1)$ has to be carefully reassessed.  In
fact what you measure when the refractive index in a given volume
of space changes is {\em not} directly the static Casimir energy
of the ``in'' state, but rather the fraction of this static Casimir
energy that is converted into photons. We have just seen that once
conversion efficiencies are taken into account, the volume dependence
is conserved, but not the power in the difference of the refractive
index~\footnote{Indeed the dependence of $|\beta|^2$ on $(n-1)^2$
and the symmetry of the former under the interchange of ``in'' and
``out'' state also proves that it is the amount of change in the
refractive index and not its ``direction'' (from ``in'' to ``out'')
that governs particle production. This also implies that any argument
using static Casimir energy balance over a full cycle has to be
used very carefully.  Actually the total change of the Casimir
energy of the bubble over a cycle would be zero (if the final
refractive index of the gas is again 1).  Nevertheless in the
dynamical calculation one gets photon production in both collapse
as well expansion phases. (Although some destructive interferences
between the photons produced in collapse and in expansion are
conceivable, these will not be really effective in depleting photon
production because of the substantial dynamical difference between
the two phases and because of the, easy to check, fact that most
of the photons created in the collapse will be far away from the
emission zone by the time the expansion photons would be created.)
This apparent paradox is easily solved by taking into account
that the main source of energy is the sound field and that the
amount of this energy actually converted in photons during each
cycle is a very tiny amount of the total power.}.

\subsection{Finite-volume numerical estimates:}

For finite volume one can no longer rely on analytic results.
Fortunately we know that for the total Casimir energy the next
subdominant term is a surface area term that is suppressed by a
factor of the cutoff wavelength divided by the bubble radius~\cite{MV}.
Canonical estimates are: $\lambda_{\mathrm cutoff}/R_{\mathrm max}
\approx 0.3 \; \mu {\rm m}/45 \; \mu {\rm m} \approx 1/150$.  This
suggests that the effects of finite bubble size will not be too
dramatic ($1\%$ in total energy?). Applying a mixture of semi-analytic
and numerical techniques~\footnote{For details, interested parties
are referred to~\cite{LBVS}.} to formula~(\ref{E-beta-squared}) we
numerically derive the spectrum $dN/d\omega$ given in Fig.~\ref{fig:thsp}.
For comparison we have also plotted the large volume analytic
approximation ({\em i.e.}, the leading bulk term by itself).

\bfg{4}
\begin{picture}(,)
\centerline{\psfig{figure=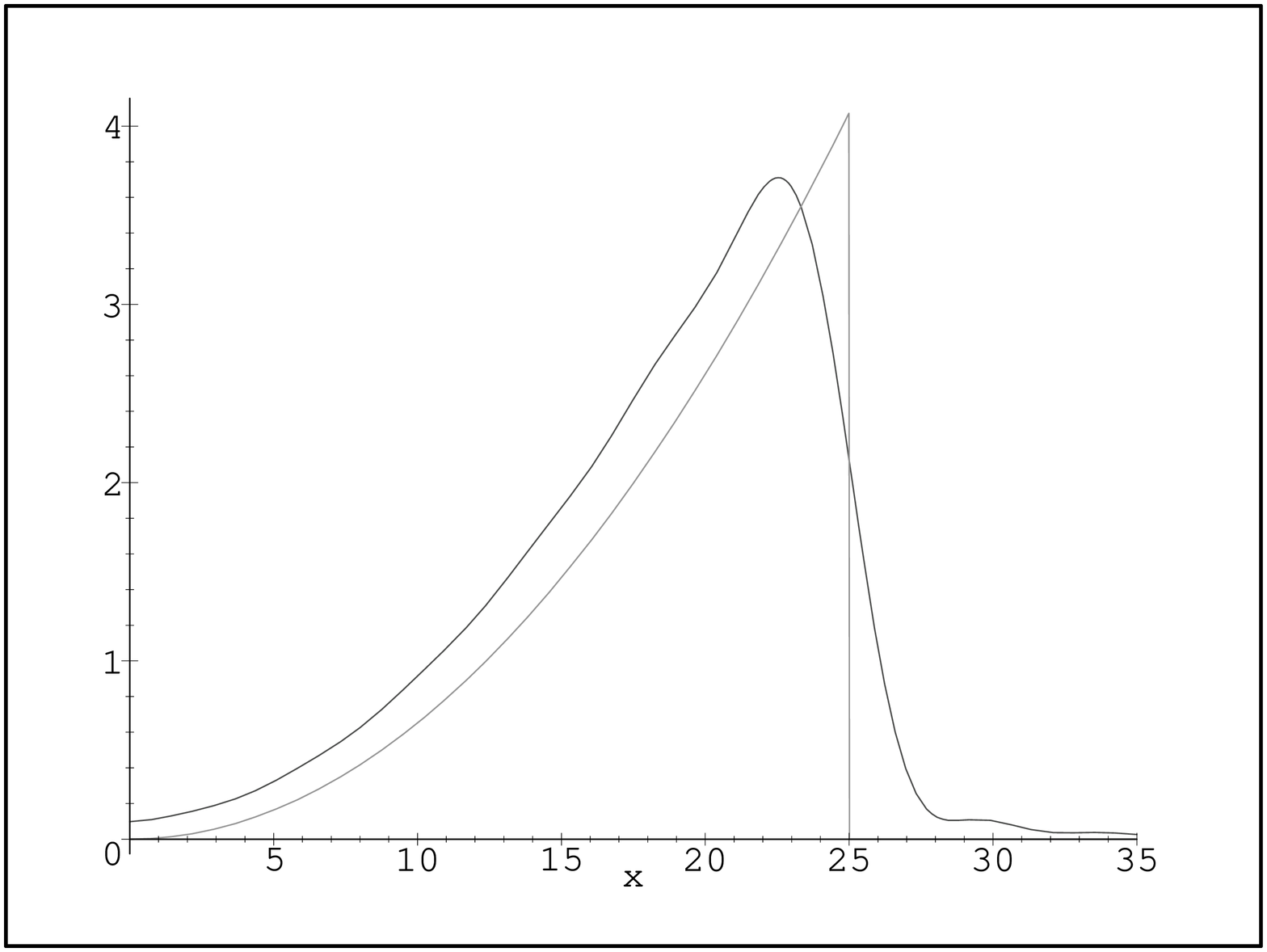,height=8cm,angle=270}}
\end{picture}
\caption{Spectrum obtained by numerical estimate for finite volume.
We have made $R\approx R_{ambient}\approx 4.5 \mu m$ and 
$\omega_{\mathrm cutoff}\approx 10^{15} Hz$. 
The sharply peaked curve is that appropriate to the (re-scaled)
infinite volume limit (Schwinger limit).}
\label{fig:thsp}
\efg

\subsection{Comment on the calculation}

The lessons we have learned from this test calculation are:

(1) The model proves (in an indirect way) that the Casimir energy
liberated via the bubble collapse includes (in the large $R$ limit) a
term proportional to the volume (actually to the volume over which the
refractive index changes). In the case of a truly dynamical model one
expects that the energy of the photons so created will be provided by
other sources of energy ({\em e.g.}, the sound wave), nevertheless the
presence of a volume contribution appears unavoidable.

(2) In spite of its simplicity (remember that the model is still
semi-static), the present calculation is already able to fit some of the 
experimental requirements, like the shape of the spectrum and the 
number of emitted photons in the case of $R=R_{\mathrm  max}$.

Of course the present model is still unable to fully fit other
experimental features. For example it provides (like the original
Schwinger model) maximal photon release at maximum expansion, and
it is able to accommodate only a few arguments to explain
the experimental dependencies. This means that a fully dynamical
calculation is required in order to deal with these issues, and it
is in order to understand what sort of model will ultimately be
required that we shall now discuss in detail some basic features
of sonoluminescence.

\section{Hints towards a truly dynamical model}

One of the key features of photon production by a space-dependent
and time-dependent refractive index is that for a change occurring
on a timescale $\tau$, the amount of photon production is exponentially
suppressed by an amount $\exp(-\omega\tau)$. In an Appendix
of~\cite{LBVS} we provided a specific toy model that exhibits this
behaviour, and argued that the result is in fact generic.

The importance for SL is that the experimental spectrum is {\em
not\,} exponentially suppressed at least out to the far ultraviolet.
Therefore any mechanism of Casimir-induced photon production based
on an adiabatic approximation is destined to failure: Since the
exponential suppression is not visible out to $\omega\approx10^{15}
\; \hbox{Hz}$, it follows that {\em if\,} SL is to be attributed
to photon production from a time-dependent refractive index ({\em
i.e.}, the dynamical Casimir effect), {\em then} the timescale for
change in the refractive index must be of order of a {\em
femtosecond}~\footnote{It would be far too naive to assume that
femtosecond changes in the refractive index lead to pulse widths limited
to the femtosecond range. There are many condensed matter processes that
can broaden the pulse width however rapidly it is generated. Indeed, the
very experiments that seek to measure the pulse width \cite{Flash1,Flash2}
also prove that when calibrated with laser pulses that are known to be of
femtosecond timescale, the SL system responds with light pulses on the
picosecond timescale.}.
Thus any Casimir--based model has to take into account that {\em
the change in the refractive index cannot be due just to the change
in the bubble radius}.

This means that one has to divorce the change in refractive index
from direct coupling to the bubble wall motion, and instead ask
for a rapid change in the refractive index of the entrained gases
as they are compressed down to their van der Waals hard core.
Yablonovitch~\cite{Yablonovitch} has emphasized that there are a
number of physical processes that can lead to significant changes
in the refractive index on a sub-picosecond timescale. In particular,
a sudden ionization of the gas compressed in the bubble would lead
to an abrupt change, from $1$ to $\approx 0$, of the dielectric
constant.

Now to get femtosecond changes in refractive index over a distance of
about 100 nm (which is the typical length scale of the emission
zone), the change in refractive index has to propagate at about $10^8$
metres/sec, about 1/3 lightspeed.  To achieve this, one has to adjust
basic aspects of the model: we feel that we must move away from
the original Schwinger suggestion, in that it is no longer the
collapse from $R_{\mathrm max}$ to $R_{\mathrm min}$ that is
important. Instead we postulate a rapid (femtosecond) change in
refractive index of the gas bubble when it hits the van der Waals hard
core~\cite{LBVS}.

We stress that this conclusion, though it moves slightly away
from the original Schwinger proposal, is still firmly
within the realm of the dynamical Casimir effect approach to
sonoluminescence. The fact is that the present work shows clearly that
a viable Casimir ``route'' to SL cannot avoid a ``fierce marriage''
between QFT and features related to condensed matter physics.

It is thus crucial to look for possible unequivocal signatures of
the dynamical Casimir effect.  To this end it is theoretically
possible to have a sharp distinction between any Casimir-like
mechanism and other proposals implying a thermal spectrum by looking
at the variance of carefully chosen two-photon
observables~\cite{BLVS}.  As a short example of how this can be
done I shall give a brief description of a way to discriminate
between thermal photons and two-mode squeezed-state photons (for
a more detailed discussion see~\cite{BLVS}).

Define the observable
\begin{equation}
N_{ab} \equiv N_{a}-N_{b},
\end{equation}
and its variance
\begin{equation}  
\Delta (N_{ab})^2=
\Delta N_{a}^{2}+\Delta N_{b}^{2}
-2 \langle N_{a} N_{b}\rangle
+2 \langle N_{a}\rangle \langle N_{b} \rangle.
\end{equation}
The number operators $N_{a},N_{b}$ are intended to denote two
photon modes, {\em e.g.} back to back photons.  In the case of true
thermal light we get
\begin{equation}
\Delta N_{a}^{2} = \langle N_{a}\rangle(\langle N_{a} \rangle +1),
\end{equation}
\begin{equation}
\langle N_{a} N_{b}\rangle = \langle N_{a}\rangle \langle N_{b}\rangle,
\end{equation}
so that
\begin{equation}  
\Delta (N_{ab})^2_{\mathrm thermal\ light}
=\langle N_{a}\rangle(\langle N_{a}\rangle+1)
+\langle N_{b}\rangle(\langle N_{b}\rangle+1).
\end{equation}
For a two-mode squeezed-state is easy to see~\cite{barnett}
\begin{equation}
\Delta (N_{ab})^2_{\mathrm two\ mode\ squeezed\ light}=0.
\end{equation}  
In fact due to correlations, $\langle N_{a} N_{b}\rangle \neq \langle
N_{a}\rangle \langle N_{b}\rangle$. Note also that, if you measure
only a single photon in the pair, you get, as expected, a thermal
variance $\Delta N_{a}^{2} = \langle N_{a}\rangle(\langle
N_{a}\rangle+1)$. Therefore a measurement of the variance $\Delta
(N_{ab})^2$ can be decisive in discriminating if the photons are
really thermal or if nonclassical correlations between the photons
occur~\cite{bk2}.

In~\cite{BLVS,letter,LBVS} it is shown that the arguments just
discussed push dynamical Casimir effect models for SL into a rather
constrained region of parameter space and predict some typical
``signatures'' for it. This allows to hope that these ideas will
become experimentally testable in the near future.

\section{Discussion and Conclusions}

The present calculation unambiguously verifies that a change of the
refractive index in a given volume of space is, as predicted by
Schwinger~\cite{Sc4,CMMV1,CMMV2,MV}, converted into real photons with
a phase space spectrum. We have also explained why such a change
must be sudden in order to fit the experimental data.  This leads
us to propose a somewhat different model of SL based on the dynamical
Casimir effect, a model focussed this time on the actual dynamics of
the refractive index (as a function of space and time) and not just of
the bubble boundary (in Schwinger's original approach the refractive
index changes only due to motion of the bubble wall).  This
proposal shares the generic points of strength attributable to the
Casimir route but it is now in principle able to implement the
required sudden change in the refractive index.

In summary, provided the sudden approximation is valid, changes in the
refractive index will lead to efficient conversion of zero point
fluctuations into real photons.  Trying to fit the details of the
observed spectrum in sonoluminescence then becomes an issue of
building a robust model of the refractive index of both the ambient
water and the entrained gases as functions of frequency, density, and
composition. Only after this prerequisite is satisfied will we be
in a position to develop a more complex dynamical model endowed with
adequate predictive power.

In light of these observations we think that one can also derive
a general conclusion about the long standing debate on the actual
value of the static Casimir energy and its relevance to
sonoluminescence: Sonoluminescence is not directly related to the {\em
static} Casimir effect.  The static Casimir energy is at best capable
of giving a crude estimate for the energy budget in SL.  We hope that
this work will convince everyone that only models dealing with the
actual mechanism of particle creation (a mechanism which must have the
general qualities discussed in this article) will be able to
eventually prove, or disprove, the pertinence of the physics of the
quantum vacuum to Sonoluminescence.

\section*{References}

\end{document}